# Temporal Variations of the Three Geomagnetic Field Components at Colaba Observatory around the Carrington Storm in 1859

Hisashi Hayakawa (1 – 4)*, Heikki Nevanlinna (5), Séan P. Blake (6 – 7), Yusuke Ebihara (8), Ankush T. Bhaskar (9), Yoshizumi Miyoshi (1).

(1) Institute for Space-Earth Environmental Research, Nagoya University, Nagoya, 4648601, Japan
(2) Institute for Advanced Research, Nagoya University, Nagoya, 4648601, Japan
(3) Science and Technology Facilities Council, RAL Space, Rutherford Appleton Laboratory, Harwell Campus, Didcot, OX11 0QX, UK
(4) Nishina Centre, Riken, Wako, 3510198, Japan
(5) Finnish Meteorological Institute, Helsinki, FI-00560, Finland
(6) Heliophysics Science Division, NASA Goddard Space Flight Center, Greenbelt, MD, USA
(7) Catholic University of America, Washington DC, United States
(8) Research Institute for Sustainable Humanosphere, Kyoto University, Uji, 6110011, Japan
(9) Space Physics Laboratory, Vikram Sarabhai Space Centre, Thiruvananthapuram, 695022, India

* hisashi@nagoya-u.jp

**Abstract**

The Carrington storm in 1859 September has been arguably identified as the greatest geomagnetic storm ever recorded. However, its exact magnitude and chronology remain controversial, while their source data have been derived from the Colaba $H$ magnetometer. Here, we have located the Colaba 1859 yearbook, containing hourly measurements and spot measurements. We have reconstructed the Colaba geomagnetic disturbances in the horizontal component ($\Delta H$), the eastward component ($\Delta Y$), and the vertical component ($\Delta Z$) around the time of the Carrington storm. On their basis, we have chronologically revised the ICME transit time as ≤ 17.1 hrs and located the $\Delta H$ peak at 06:20 – 06:25 UT, revealing a magnitude discrepancy between the hourly and spot measurements (−1691 nT *vs*. −1263 nT). Furthermore, we have newly derived the time series of $\Delta Y$ and $\Delta Z$, which peaked at $\Delta Y \approx$ 378 nT (05:50 UT) and 377 nT (06:25 UT), and $\Delta Z \approx$ −173 nT (06:40 UT). We have also computed the hourly averages and removed the solar quiet ($Sq$) field variations from each geomagnetic component to derive their hourly variations with latitudinal weighting. Our calculations have resulted in the disturbance variations (*Dist*) with latitudinal weighting of *Dist* $Y \approx$ 328 nT and *Dist Z* ≈ −36 nT, and three scenarios of *Dist H* ≈ −918, −979, and −949 nT, which also approximate the minimum Dst. These data may suggest preconditioning of the geomagnetic field after the August storm (Δ*H* ≤ −570 nT), which made the September storm even more geoeffective.

**1. Introduction**

Solar eruptions occasionally launch geo-effective interplanetary coronal mass ejections (ICMEs), which cause geomagnetic storms and extend the auroral oval equatorward (Gonzalez *et al*., 1994; Daglis *et al*., 1999; Hudson, 2021). Analyses of such space weather events are important not only for improving our knowledge of the solar-terrestrial environment, but also for assessing the social impact of space weather, as modern civilisation has become increasingly vulnerable to extreme space weather events through its increasing dependence on technological infrastructure (Baker *et al*., 2008; Lanzerotti, 2017; Riley *et al*., 2018; Hapgood *et al*., 2021). Among recorded space weather events, the Carrington storm on 1859 September 2 is frequently described as a worst-case scenario, in terms of the impact that such an extreme geomagnetic disturbance (Tsurutani *et al*., 2003; Siscoe *et al*., 2006; Cliver and Dietrich, 2013) would have on modern infrastructure (Baker *et al*., 2008; Riley *et al*., 2018; Oughton *et al*., 2019; Hapgood *et al*., 2021).

The Carrington storm forms one of the benchmarks in space weather studies. It is associated with the earliest reported white-light flare on 1859 September 1 (Carrington, 1859; Hodgson, 1859) and one of the most intense flares, fastest ICMEs, geomagnetic disturbances, and auroral extensions in the observational history (Tsurutani *et al*., 2003; Cliver and Svalgaard, 2004; Boteler, 2006; Green and Boardsen, 2006; Silverman, 2006; Cliver and Dietrich, 2013: Freed and Russell, 2014; Curto *et al*., 2016; Hayakawa *et al*., 2019, 2020; Miyake *et al*., 2019). Its geomagnetic disturbance has been variously estimated for minimum Dst index of ≈ −1760 nT in spot values and ≈ −850 to −1050 nT in hourly averages, according to the Colaba *H* magnetometer (Tsurutani *et al*., 2003; Siscoe *et al*., 2006; Gonzalez *et al*., 2011; Cliver and Dietrich, 2013). This magnetometer also captured an exceptionally intense negative Δ*H* excursion of ≈ −1600 nT (fig. 3 of Tsurutani *et al*., 2003; fig. 1a of Kumar *et al*., 2015). In the mid-19th century, British colonial observatories conducted magnetic measurements in mainland England, Ireland, Canada, Australia, India, and South Africa. Among them, the Colaba Observatory managed to obtain a unique record of this storm in 15-min cadence in the stormy interval and hourly cadence otherwise, allegedly without data gaps, in the low to mid magnetic latitudes (MLATs) (Tsurutani *et al*., 2003). The Colaba records are contrasted with other magnetograms from mid to high MLATs, which were most likely affected by auroral electrojets and field-aligned currents (Nevanlinna, 2006, 2008; Blake *et al*., 2020).

However, interpretation of this geomagnetic superstorm has been challenging. This exceptionally large negative excursion has been controversially explained by an enhancement of the ring current (Tsurutani *et al*., 2003; Keika *et al*., 2015), auroral electrojet (Akasofu and Kamide, 2004; Green and Boardsen, 2006; Cliver and Dietrich, 2013), and field-aligned currents (Cid *et al*., 2015). The Colaba *H* dataset has been subjected to numerous geospace simulations by considering balance between solar wind energy input and loss of ring current ions (Keika *et al*., 2015; Blake *et al.*, 2021). The time series of the storm has also been the subject of some controversies, as the peak magnitude has been located at either 10:26 (fig. 3 of Tsurutani *et al*., 2003) or 11:12 (fig. 1a of Kumar *et al*., 2015) in Bombay local time (LT). Furthermore, the contemporary solar quiet (*Sq*) field variations have not been evaluated, whereas — by definition — these variations must be subtracted from the Δ*H* time series when reconstructing Dst index (Sugiura, 1964; Yamazaki and Maute, 2017). In this context, we have recently located a published version of the Colaba yearbook for 1859 (Fergusson, 1860), containing source tables for geomagnetic measurements of the horizontal force (*H*) component, as well as the eastern declination (*D*) and vertical force (*Z*) components (Figure 1). On this basis, we modified the controversial magnitude and time series for the Colaba *H* component, newly derived the Colaba *D* and *Z* components around the Carrington storm, and assessed the impact of contemporary *Sq* variations to form a quantitative basis for further scientific discussions of the Carrington storm.

## 2. Materials and Methods

The Colaba Observatory was situated in Bombay (N18°54', E072°48') and had conducted magnetic measurements since 1845. In 1859, the Colaba Observatory measured geomagnetic variations with declinometers, two horizontal force magnetometers (large and small), and one vertical force magnetometer with instrumental thermometers, dip circles, and apparatus for deflection (Fergusson, 1860, pp. vi – xiii). From 1846–1847, the observatory continued using Grubb's *large* magnetometers (Royal Society, 1840) and supplemented these measurements with *small* magnetometers (unifilar and bifilar portable magnetometers; see Riddell, 1842; Tsurutani *et al*., 2003). The deflection apparatus was used to determine absolute *H*, approximately every week.

[Figure 1 image: tables from Bombay Magnetical Observations, pp. 83 & 169]

figure 1: Excepts from the Colaba yearbook, showing the hourly-value tables for 1–4 September 1859 and spot-value tables for 1–2 September 1859 (Fergusson, 1860, pp. 83 & 169).

Regular magnetic and meteorological observations at Colaba Observatory were recorded in their archives and published in yearbooks. Copies of the Colaba 1859 yearbook (Fergusson, 1860) can be found in several archives such as the India Office Records and Private Papers of the British Library (IOR/V/18/215). This yearbook contained tabulated geomagnetic measurements of the eastern $D$, $H$, and $Z$ components, with astronomical timestamps (running from noon to noon) in Göttingen Mean Time (GöMT = UT + 40 min − 12 hrs). These measurements have been summarised in two series of tables (Figure 1). The hourly tables do not provide hourly averages, but rather hourly spot measurements conducted regularly, except on Sundays and certain holidays (Fergusson, 1860, pp. vii and 2 – 153). Additionally, spot values of 'disturbance observations' were recorded every 15 minutes — and occasionally every 5 to 10 minutes — during significant geomagnetic disturbances (Fergusson, 1860, p. vii). The latter data offer slightly more detailed data for the stormy interval than in Tsurutani et al. (2003), which visualised the data in 15-min candence during the stormy interval. The large magnetometer measured the spot values of the $D$ component at full time, the $H$ component 2 min after full time, and the *Z* component 2 min before full time (Fergusson, 1860, pp. 154 – 179).

The hourly measurement tables from Grubb's *large* magnetometers record the eastern *D* measurements in angular minutes (′), while the *H* and *Z* measurements are recorded in absolute values with English Units (EU) and scale readings with temperature corrections (Figure 1a), where 1 EU equals 4610.8 nT (Barraclough, 1978, p. 3). The spot-measurement tables for the *large* and *small* magnetometers commonly present the eastern *D* measurements in angular minutes but the *H* and *Z* measurements only as scale readings, without temperature measurements, while the instrumental temperature ($T(t)$) is presented separately in °F (Figure 1b).

From the records in this yearbook, we have derived the variations in $\Delta Y$, $\Delta H$, and $\Delta Z$ at Colaba Observatory in 1859. We first derived the baselines of the three reported components ($D_B$, $H_B$, and $Z_B$), selecting the five quiet days in 1859 August based on the Ak index (Nevanlinna, 2004) and averaging their absolute measurements on the closest quiet day to the storm onset (August 25 in civil GöMT). Following contemporary textbooks (Gauss, 1838; Lamont, 1867), we derived the $\Delta Y$ variations using Equation (1), abbreviating the reported *D* variables as $D(t)$. Our approximation is valid for $D(t) - D_B = \Delta D(t) \ll 1°$, which was actually the case at Colaba at that time (Figure 1). Here, we need to emphasise that the *H* and *Y* components are not orthogonal. Still, northward component ($\Delta X$) approximates with $\Delta H$ here, as the eastern declination remained $\ll 1°$ (Figure 1; Fergusson, 1860).

$$\Delta Y(t) = H_B \{\sin(D(t)) - \sin(D_B)\} \approx H_B (D(t) - D_B) \ldots (1)$$

The hourly *H* values are provided as both absolute values ($H_{AB}$), in EU, and as scale reading values ($H_{SR}$), as shown in Figure 1a. Their values are based on the *large H* magnetometer, as the *small* magnetometer was only used as a crosscheck "under various disadvantages" (Fergusson, 1860, p. xi). The yearbook (Fergusson, 1860, p. x) uses Equation (2) to describe the relationship between $H_{AB}$ and $H_{SR}$, where *T* represents the temperature of the thermometer (in °F) attached to the large horizontal magnetometer. The hourly tables verify this equation with a steady offset of $H_{AB} = H_{SR} + 28 \pm 1$. In our study, we converted these parameters to the modern unit (nT) and corrected this steady drift, as summarised in Equation (3). Here, the most relevant coefficients are the sensitivity, for converting *H*-scale values into nanoTeslas (75.62 nT/scale division), and the temperature coefficient of 13.6 nT for each degree of Fahrenheit. For Z, the respective coefficients are 50.72 nT/scale div and 1.5 nT per Fahrenheit. The temperature coefficients seem slightly large in *H* and slightly small in *Z*. Their causes may be better understood if we can in future locate and analyse their original magnetometers used in Colaba at that time. The *H* baseline ($H_B$) was subtracted when deriving $\Delta H$ variations (Equation (4)).

$$H_{AB}(t) \text{ [EU]} = 8.0340 + 0.0164 \{H_{SR}(t) + 0.18(T-80) - 20.00\} \ ... \ (2)$$
$$H_{AB}(t) \text{ [nT]} = 4610.8 [8.0340 + 0.0164 \{H_{SR}(t) + 0.18(t-80) - 20.00\}] + 28 \pm 1 \ ... \ (3)$$
$$\Delta H(t) = H_{AB}(t) - H_B \ ... \ (4)$$

In the hourly *Z* table, Fergusson (1860) provided two columns for the *Z* measurements, as scale readings ($Z_{SR}(t)$) and absolute values ($Z_{AB}(t)$), where $Z_{AB}$ was calculated from the absolute *H* and *I*, $Z_{AB} = H_{AB} \tan(I_A)$. While the conversion equation is not clarified in the 1859 yearbook, the 1860 yearbook (Fergusson, 1861, p. xiii) allows us to summarise it as Equation (5). In 1860, the contemporaneous baseline (*Q*) varied over time, with values of 2.78821 from January 1 to October 9, 2.8652 from October 9 to December 29, and 3.0491 after December 29. If we apply the initial value (*Q* = 2.78821), this shows a steady offset of $H_{AB} = H_{SR} - 504 \pm 2$, which was corrected using Equation (6). We derived $\Delta Z$ taking the *Z* baseline ($Z_B$) into account (Equation (7)).

$$Z_{AB} \text{ [EU]} = Q + 0.011 \{Z_{SR} + 0.03(T-80) - 40.0\} \ ... \ (5)$$
$$Z_{AB} \text{ [nT]} = 4610.8 [Q + 0.011 \{Z_{SR} + 0.03(T-80) - 40.0\}] - 504 \pm 2 \ ... \ (6)$$
$$\Delta Z(t) = Z_{AB}(t) - Z_B \ ... \ (7)$$

3. Results

Figure 2 illustrates our reconstruction of the geomagnetic measurements of $\Delta H$, $\Delta Y$, and $\Delta Z$ at the Colaba Observatory from 1859 August 26 to September 5, with the timestamps corrected from GöMT to UT. This figure shows two extreme geomagnetic storms on August 28/29 and September 2. This figure only shows the recovery phase of the August storm, as observations were not conducted on August 28 because it was on Sunday (Fergusson, 1860, pp. vii). The intensity of these measurements for 28 August can be conservatively interpreted as $\Delta H \leq -570$ nT, $\Delta Y \geq 55$ nT, and $\Delta Z \geq 128$ nT, respectively. Following the August storm, the geomagnetic field intensities recovered to only $\Delta H \approx -85$ nT, $\Delta Y \approx 9$ nT, and $\Delta Z \approx 77$ nT (at local midnight on 1/2 September), respectively.

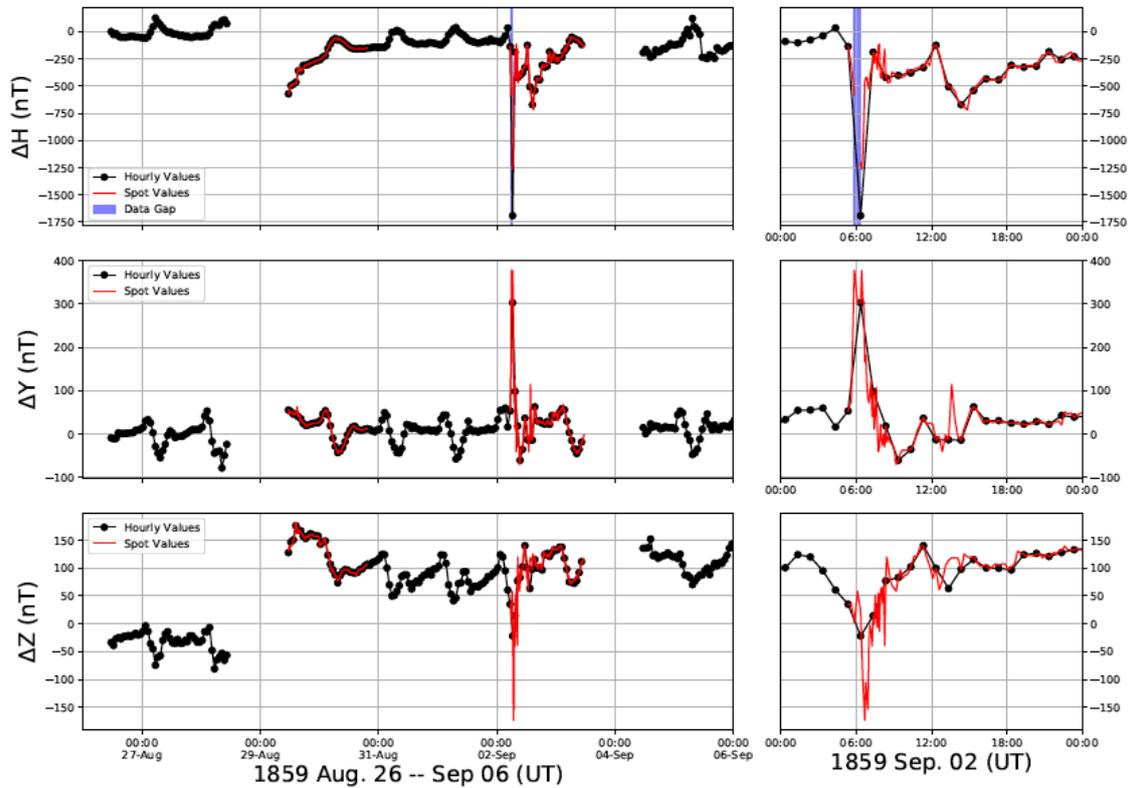

Figure 2: Spot values[1] (red) and hourly values[2] (black) of ΔH, ΔY, and ΔZ at Colaba Observatory indicating geomagnetic disturbances, as reconstructed from the Colaba Yearbook (Fergusson, 1860). These hourly values are not the hourly averages but hourly spot measurements. The ΔH data gap range is shown in blue.

The September storm started at 04:50 UT, according to Bartels (1937), whereas the storm commencement (SC) peaked slightly earlier at 04:20 UT (17:00 GöMT), as shown in Figure 2. This indicates an ICME transit time of ≤ 17.1 hrs (*vs*. 17.6 hrs in Freed and Russell (2014)) and an SC amplitude of ≥ 119 nT (*vs*. ≥ 120 nT in Tsurutani *et al.* (2003) and ≈ 113 nT in Siscoe *et al.* (2016)), taking the chronological offset with the reported solar flare onset at 11:15 UT (Carrington, 1859) and the intensity offset with the pre-storm level at local midnight (19:20 UT) into consideration, respectively. These values are no more than conservative estimates, as they are derived from the hourly spot values, which may have missed the actual SC onset and the actual SC peak.

The storm developed rapidly after the SC peak at 04:20 UT. The geomagnetic field intensities

---

[1] https://www.kwasan.kyoto-u.ac.jp/~hayakawa/data/Carrington_Colaba/SD1_1859_CLA_spot.txt
[2] https://www.kwasan.kyoto-u.ac.jp/~hayakawa/data/Carrington_Colaba/SD2_1859_CLA_hourly.txt

peaked at Δ*H* ≈ −1263 nT (06:25 UT = 19:05 GöMT), Δ*Y* ≈ 378 nT (05:50 UT = 18:30 GöMT) and 377 nT (06:25 UT = 19:05 GöMT), and Δ*Z* ≈ −173 nT (06:40 UT = 19:20 GöMT). Our Δ*H* time series chronologically supports the findings of Kumar *et al*. (2015) over those of Tsurutani *et al*. (2003), who located the Δ*H* peak at 06:20 UT (11:12 in Bombay LT) and 05:34 (10:26 Bombay LT), respectively. However, several caveats must be noted here. Firstly, the pre-storm level was slightly different from the initial baseline, as shown in this section. Secondly, we detected a data gap in the *H* measurement at 06:05 UT (18:45 GöMT). Finally, and most importantly, the spot Δ*H* amplitude (−1263 nT) departs from the hourly Δ*H* amplitude of ≈ −1691 nT at 06:20 UT (19:00 GöMT), whereas the hourly values of Δ*Y* (303 nT at 06:20 UT) and Δ*Z* (−22 nT at 06:20 UT) are more moderate.

The *H* error margin was described as 0.008 EU (= 37 nT) in Fergusson (1860, p. xi). We have further computed the Δ*Y* error margins as 11 nT or 22 nT, following Equation 1 and assuming the D reading accuracy as 1′ or 2′, respectively. The Δ*Z* error margins are estimated as 25 nT or 39 nT, if we assume the reading accuracy of the dip circle measurements as 1′ or 2′ and the *I* ≈ 20°. On their basis, their error margins are estimated ≈ 20 – 40 nT during the regular measurements. These estimates are valid for quiet period of the magnetic field before and after the Carrington peak. When the magnetic field is changing rapidly, like during the Carrington storm, the light spot from the mirror attached on the magnet moves on the scale quickly, and this causes problems to the observer to fix the position of the spot on the scheduled time (full time). This is probably a major source of error for the magnetic measurements during the storm. Therefore, it is extremely difficult to quantitatively calculate the error margins during the storm peak, whereas they may have reached ≈ 100 nT or even more.

**4. Storm Intensities**

Figure 2 shows a much more moderate spot Δ*H* amplitude at the Colaba Observatory (≈ −1263 nT) than in the previous estimates of ≈ −1600 nT (Tsurutani *et al*., 2003; Kumar *et al*., 2015). In contrast, the reported hourly Δ*H* amplitude (≈ −1691 nT at 06:20 UT) seems consistent with these previous estimates when we derive the baseline at local midnight immediately before the September storm (≈ −1606 nT). This hourly Δ*H* value is the only similar figure in the tables of hourly and spot values (Figure 1; Fergusson, 1860), as the spot Δ*H* value at 06:20 UT (19:00 GöMT) is ≈ −1208 nT and even milder than the spot Δ*H* value (≈ −1263 nT) at 06:25 UT (19:05 GöMT).

There are several possible explanations for this inconsistency. If we assume the original table entirely correct, this large jump can be attributed to the 2-min time lag between the measurements of hourly values and spot values (Figure 1a). This hypothesis requires an extremely sharp positive excursion of ≈ 483 nT within these 2 min (≈ 241.5 nT/min). For the rapid Dst decrease to have been caused by the ring-current development requires at least ≈ 2700 mV/m of solar wind electric field ($VB_z$), where $V$ is the solar wind speed and $B_z$ is the Z component of the interplanetary magnetic field, according to the empirical Dst model (Burton *et al.*, 1975). The solar wind electric field is usually on the order of 1 mV/m and is thought to have increased to ≈ 340 mV/m during the Carrington storm (Tsurutani and Lakhina, 2014). Thus, the ring current is unlikely to have caused an extremely sharp positive excursion of ≈ 483 nT within 2 min. Alternatively, if we critically reconsider the original table and modify the tabulated scale reading value of −244 at 06:20 UT (19:00 GöMT) to 244 (removing the minus sign), this Δ$H$ value could be modified to −1322 nT. This value is much closer to the spot values around this peak, whereas this is no more than a speculation. Here, we conservatively place caveats on the reliability of using the hourly Δ$H$ value as a spot measurement at 06:20 UT, which probably formed the basis of the greatest Δ$H$ spike in existing studies (Tsurutani *et al.*, 2003; Siscoe *et al.*, 2006; Cliver and Dietrich, 2013; Kumar *et al.*, 2015).

The Colaba magnetogram was used to estimate the Dst time series. By definition, the Dst index is derived by averaging the hourly disturbance variations (*Dist*) of the four mid/low-latitude reference stations with latitudinal weighting (Sugiura, 1962). In 1859, the Colaba Observatory was located at λ = 10.2° MLAT, according to the GUFM1 model (Jackson *et al.*, 2000). Here, we have approximated the Dst time series with the Colaba $H$ magnetometer using Equation (8).

*Dist H* (t) ≈ ($H_{AB}$ (t) – $H_B$ – *Sq* (t))/cosλ. (8)

We approximated *Sq* (t) following a classic *Sq* definition to take an average of five quietest days of a month (Chapman and Bartels, 1940, p. 214), whereas we have more modern approaches to compute *Sq* for a given time and location (*e.g.*, Van der Kamps, 2013). Here, we have selected the five quiest days in August 1859, following the Ak index (Nevanlinna, 2004). The Colaba magnetometers captured three days of their diurnal variations completely, as two of the five quietest days in August were holidays, and the records were therefore incomplete (Fergusson, 1860). Therefore, we have used the diurnal variations for these three days with complete measurements to approximate *Sq* (t) in August 1859. To remove the *Sq* variations, we followed the same procedures for the Δ$Y$ and Δ$Z$ time series, as shown in Figure 3.

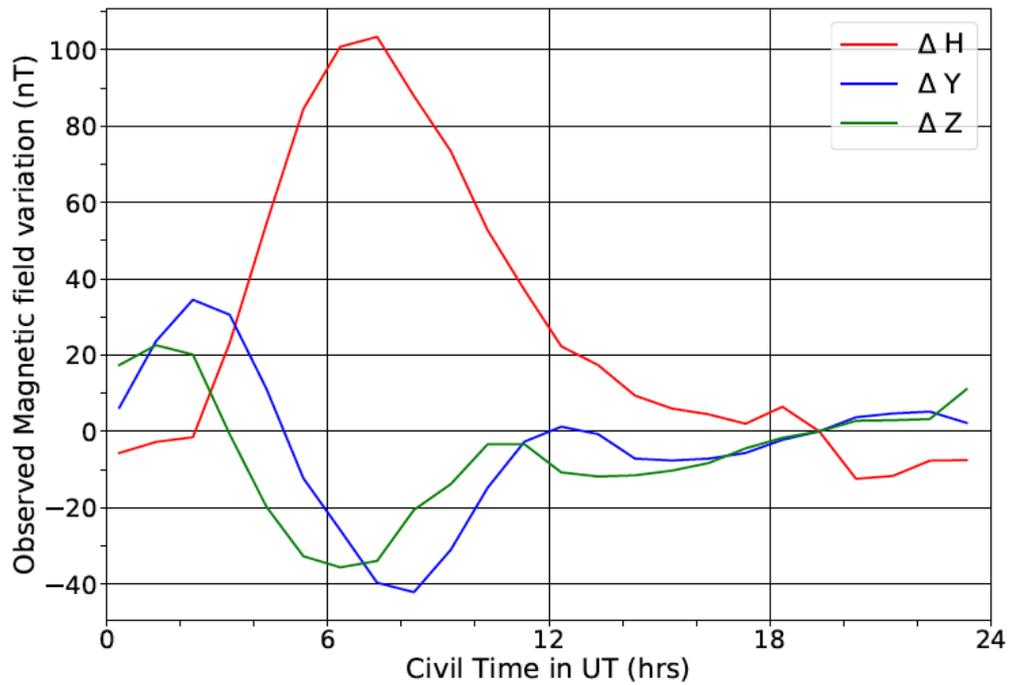

Figure 3: The solar quiet (Sq) variations of $\Delta H$ (red), $\Delta Y$ (blue), and $\Delta Z$ (green)[3], as computed from the three quietest days with complete hourly datasets.

---

[3] https://www.kwasan.kyoto-u.ac.jp/~hayakawa/data/Carrington_Colaba/SD3_1859_CLA_Sq.txt

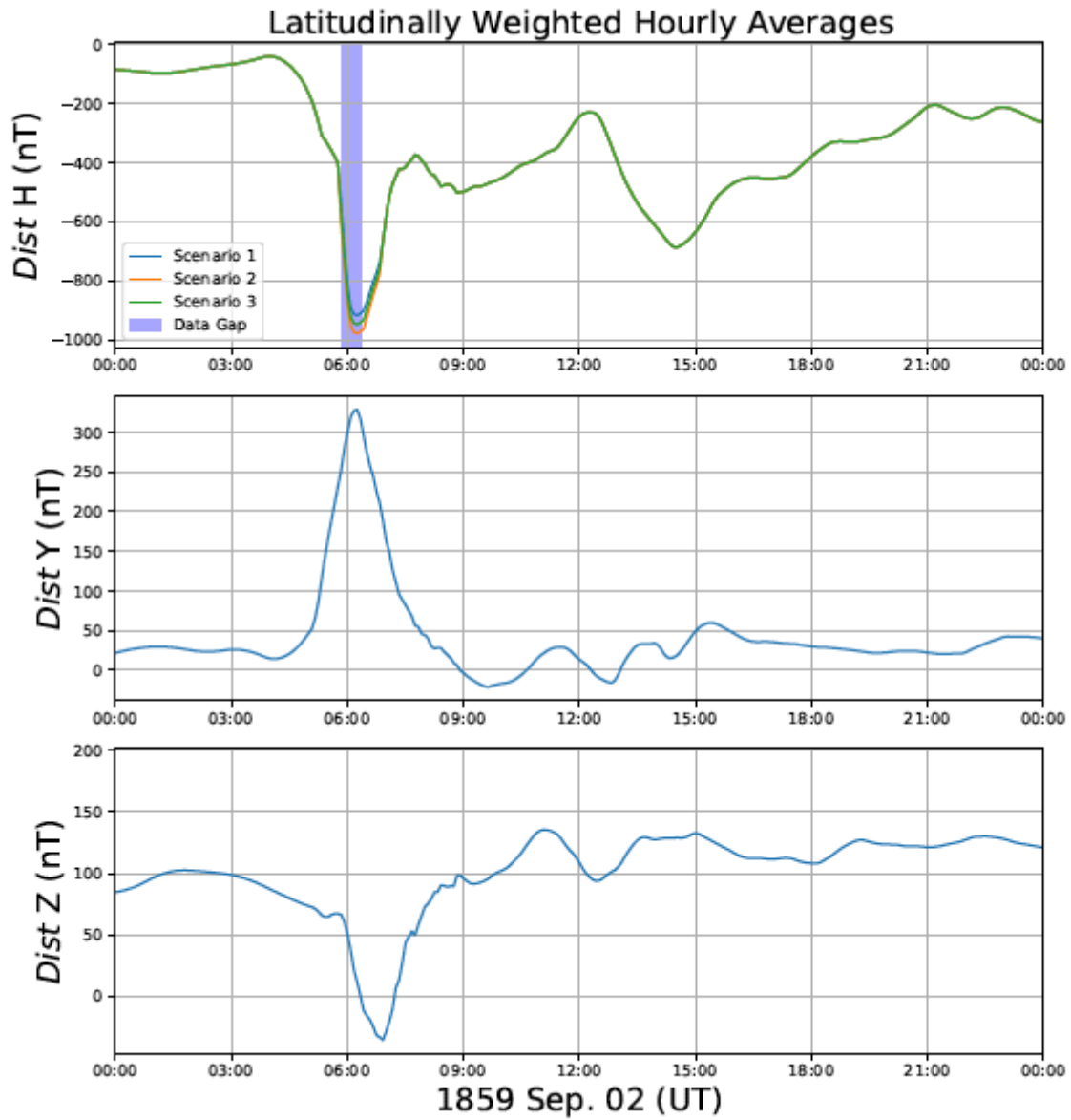

Figure 4: The latitudinally weighted hourly *Dist H*, *Dist Y*, and *Dist Z* at Colaba Observatory, after removal of their Sq variations. The *Dist H* data gap is shown in blue.

Figure 4 summarises the hourly *Dist H*, *Dist Y*, and *Dist Z* with latitudinal weighting. Here, we have interpolated the spot values to 5-min intervals and taken their hourly averages, as the intervals of the Colaba measurements were uneven around the storm peak (Figure 1). Specifically, we have plotted three scenarios for determining the *Dist H* storm peak: (1) accepting the unchanged hourly Δ*H* value at 06:20 UT (19:00 GöMT); (2) accepting only the spot Δ*H* value at 06:20 UT; and (3) taking an average of the hourly and spot Δ*H* values at 06:20 UT.

As shown in Figure 4, the geomagnetic disturbances peaked at *Dist Y* = 328 nT at 06:05 UT and *Dist Z* = −36 nT at 06:10 UT, and *Dist H* = -918 nT (Scenario 1), -979 nT (Scenario 2), and -949 nT (Scenario 3), with latitudinal weighting. The *Dist H* intensity is a conservative value, as we have a data gap at 06:05 UT (18:45 GöMT). The minimum *Dist H* roughly approximates the minimum Dst* estimate for the Carrington storm, whereas we need to be cautious on the local time effects and ultimately average this with *Dist H* in three more reference mid/low-latitude magnetometers (*e.g.*, Sugiura, 1962).

Figure 4 also shows that the September pre-storm levels of *Dist H*, *Dist Y* and *Dist Z* were different to the baselines, by ≈ −86 nT, ≈ 9 nT, and ≈ 78 nT, respectively. Accordingly, during the September storm, the magnetic field had not completely recovered from the August storm, making the September storm more effective in *Dist H* and *Dist Y* and less effective in *Dist Z*. It is slightly challenging to understand their cause, while we can still suggest several possibilities. Firstly, after the August storm, the ring current decay may have required a longer time. This scenario is unlikely, as the ring current development down to the geocorona also enhances the decay rate as well. Secondly, this jump was caused by ions with higher energy. This scenario may be possible, as higher ion energy requires longer time for the ring-current decay compared with the typical tens keV energy range (*e.g.*, Ebihara and Ejiri, 2003). Thirdly, there may have been a continuous supply of source ions for the ring current enhancement associated with substorm injections. This is also possible, if the coronal hole supplies high-speed solar wind and causes multiple substorms (Tsurutani *et al*., 2006). Furthermore, it is also known that the continuous magnetic reconnection between the southward component of the Alfvén waves and the Earth's magnetosphere fields slowly injects solar wind energy into the magnetosphere, which causes slow decay of ring current and thus the extended recoveries of the geomagnetic storms (Tsurutani *et al*., 1995, Raghav *et al*., 2018).

**5. Summary and Discussions**

In this article, we have reconstructed the geomagnetic disturbances in Δ*H*, Δ*Y*, and Δ*Z*, based on data in the recently discovered Colaba yearbook (Fergusson, 1860). Until this point, the Colaba *H* magnetometer represented the ground truth for the Carrington storm and any scientific discussions on this event since Tsurutani *et al*. (2003). However, our analyses have not only revised the Δ*H* disturbance but also derived the Δ*Y* and Δ*Z* disturbances. As shown in Figure 1, the Colaba 1859 yearbook provides two series of geomagnetic measurements, namely regular hourly measurements and intermittent spot measurements (every 5 – 15 min) during specific geomagnetic disturbances.

We converted the tabulated geomagnetic disturbances from scale readings to SI units (nT) and reconstructed their time series (Figure 2). Accordingly, we have resolved the controversial $\Delta H$ chronology and located the SC peak at 04:20 UT and the storm peak at 06:20 – 06:25 UT. This indicates that the Carrington ICME had a slightly shorter transit time than previously considered ($\leq$ 17.1 h). This yields a slightly faster average ICME velocity of $\geq$ 2430 km/s, which is slightly faster than what has been considered. We have also identified a previously unrecognised data gap at 06:05 UT and an apparent discrepancy between the hourly and spot values in the $\Delta H$ tabulations (−1263 nT *vs*. −1691 nT). This appears to be slightly abnormal, as the hourly value becomes even larger than the spot values, in contrast with what would be expected for the historical magnetograms. In addition, we have newly derived a $\Delta Y$ and $\Delta Z$ time series, which peaked at $\Delta Y \approx$ 378 nT (05:50 UT) and 377 nT (06:25 UT), and $\Delta Z \approx$ −173 nT (06:40 UT).

Our results place caveats on the existing Dst estimate for the Carrington storm, owing to the controversial $\Delta H$ peaks in the spot and hourly values. Furthermore, the definition of the Dst index requires the removal of the Sq variation and baseline, and uses the hourly average of these parameters with latitudinal weighting. Therefore, we derived the Sq variations in each component from the quiet-day measurements (Figure 3) and removed them from the reconstructed geomagnetic disturbances in each component to derive their hourly averages with latitudinal weighting (Figure 4). Accordingly, their intensities are estimated as hourly *Dist Y* = 328 nT, *Dist Z* = −36 nT, and *Dist H* = −918 nT (Scenario 1), −979 nT (Scenario 2), and −949 nT (Scenario 3). The minimum *Dist H* roughly approximates the Dst estimate for the Carrington storm, whereas the local time effect still leaves large uncertainty.

The positive $\Delta Y$ value indicates an eastward deflection of the geomagnetic field, which was probably caused by the ionospheric current flowing towards the equator. The equatorward current is thought to be part of the DP 2 ionospheric current system and two-cell magnetospheric convection (Nishida, 1968). The large amplitude of $\Delta Y$ suggests an intensification of the magnetospheric convection that is needed to transport hot plasmas and intensify the ring current (Tsurutani *et al*., 2003).

The August storm was incompletely captured in this dataset, due to the weekend break in observations. We have conservatively estimated its intensity as $\Delta H \leq$ −570 nT, $\Delta Y \geq$ 55 nT, and $\Delta Z \geq$ 132 nT. The magnetic field had not completely recovered from the August storm when the outbreak of the September storm began. This emphasises the role of the preceding August storm, which preconditioned the magnetic field and made the Carrington storm being more effective. Overall, the Colaba 1859 yearbook (Fergusson, 1860) has significantly benefitted our understanding on the space weather variations around the Carrington storm. It is worth investigating Colaba archival manuscripts to further improve our reconstructions for the Carrington storm.

**Acknowledgments**

We thank Naro Balcrushna, Ramchund Pandoorung, and Luxumon Moreshwar for their manual geomagnetic measurements conducted at Colaba Observatory in shift work during the Carrington storm. Their industrious measurements have formed irreplaceable datasets for scientific discussions of the Carrington storm. We thank the British Library for allowing us to access their collections. HH has benefited from discussions within the ISSI International Team #510 (SEESUP Solar Extreme Events: Setting Up a Paradigm) and ISWAT-COSPAR S1-01 and S1-02 teams. HH thanks Denny M. Oliveira for his helpful comments. This work was financially supported in part by JSPS Grant-in-Aids JP20K22367, JP20K20918, JP20H05643, and JP21K13957, JSPS Overseas Challenge Program for Young Researchers, and the ISEE director's leadership fund for FY2021 and Young Leader Cultivation (YLC) program of Nagoya University.